\documentclass[twocolumn,a4paper,aps,pr,floatfix]{revtex4}
\usepackage{amssymb,amsmath,textcomp,epsfig,txfonts,subfigure}
\usepackage{graphicx}

\newcommand{\comment}[1]{}

\begin{document}
\preprint{$ $Id: echotech.tex,v 1.32 2005/06/08 09:44:36 propan Exp $ $}
\draft
\title{Multi-speckle diffusing wave spectroscopy with a single
mode detection scheme }
\author{P. Zakharov, F. Cardinaux and F. Scheffold }
\address{$^{(1)}$Department of Physics, University of Fribourg,
CH-1700 Fribourg, Switzerland; email: Frank.Scheffold@unifr.ch}
\date{\today}

\begin{abstract}
We present a  detection scheme for diffusing wave spectroscopy
(DWS) based on a two cell geometry that allows efficient ensemble
averaging. This is achieved by putting a fast rotating diffuser in
the optical path between laser and sample. We show that the
recorded (multi-speckle) correlation echoes provide an ensemble
averaged signal that does not require additional time averaging.
Furthermore, combined with traditional two-cell DWS, the full
intensity autocorrelation function can be measured with a single
experimental setup. The new scheme provides access to a large
range of correlation times thus opening a new experimental window
for the study of slowly relaxing and arrested systems, such as
viscoelastic complex fluids, colloidal glasses and gels.
\end{abstract}

\maketitle

The surging interest in slowly relaxing and arrested colloidal
systems such as gels or glasses
\cite{cipelletti02:slow,pham02:multiple,eckert02:glass} has
created a need to monitor dynamic properties on time scales of
seconds and minutes. Light scattering is certainly one of the best
methods for this purpose since it offers convenient access to such
key dynamic properties as the intermediate scattering function or
the particle mean square displacement. Traditionally a single
speckle mode of scattered light is detected and fluctuations are
recorded over a time much longer that the relaxation time.
However, this time averaging scheme is not applicable to rigid,
nonergodic systems. For these systems the ensemble average can be
obtained by summing a collection of consecutive experiments
conducted on different sample realizations. Usually the sample is
translated or rotated and a scan over a large number of
independent speckles is performed
\cite{pusey89:dls,scheffold03:light,xue92:nonergodicity,schatzel93:accuracy}.
A major drawback of this approach is the extensive duration of
measurements. It is not unusual today to investigate relaxation
process on time scales of seconds and minutes with a corresponding
measurement time of hours and days. As a matter of fact several
authors have reported data collection time more than a day for a
single intensity correlation function (e.g. \cite{beck99:glass,
megen94:glass}). Besides being tedious and time consuming this
approach is restricted to the systems in (quasi-)equilibrium. Only
the advent of multi-speckle detection schemes made it possible to
conveniently monitor very slow relaxation processes. Dynamic light
scattering using a digital (CCD/CMOS) camera as a detector offers
the possibility to perform simultaneously a large number of
independent experiments thus achieving ensemble averages in real
time \cite{bartsch97:glass}. Compared to single photon counting
the limited dynamic range of digital cameras and the typically
high dark counts result in a somewhat lower accuracy. But
unfortunately with a time resolution of typically 1-10~ms digital
camera based detection is restricted to rather long correlation
times. Thus, traditional photon correlation spectroscopy has to be
made as well in a separate experiment if access to the full range
of correlation times is required.

 All considerations above apply equally to both dynamic light
scattering (DLS) in the single scattering regime and diffusing
wave spectroscopy (DWS) in the multiple scattering regime
\cite{berne76,maret1987,pine1988}. However, due to the strong
multiple scattering DWS offers more flexibility in the
experimental design, which we have exploited in our approach. In
this paper we report on a  new two-cell detection scheme for
diffusing wave spectroscopy (DWS) that provides an effective
multi-speckle averaging using single mode detection. To obtain an
ensemble averaged signal we illuminate our sample with the laser
light scattered from a rotating diffuser and we analyze reflected
or transmitted light. We show that echoes in the recorded
correlation function appear at any revolution while the
correlation function of the sample remains finite. Each echo
signal is generated by a large number of independent speckles thus
efficient ensemble averaging is performed. Moreover, we
demonstrate that the intensity correlation function of the sample
can be extracted from the two-cell echoes.

The detection of single and multiple scattering correlation echoes
was discussed in previous articles
\cite{hebraud97:yielding,hohler97:periodic,megen98:measurement,pham04:echo,petekidis02:rearrangements}.
Echo DWS was initially introduced in the analysis of non-linear
shear deformation \cite{hebraud97:yielding,hohler97:periodic}. In
the single scattering regime Pham et al. recently demonstrated the
use of echo DLS for efficient ensemble averaging
\cite{pham04:echo}. Our DWS echo-scheme implements a new physical
optical principle to record the ensemble averaged intensity
correlation function. In contrast to previous echo experiments in
our case the sample is at rest. Thus high rotation or oscillation
frequencies can be realized without any mechanical disturbance of
the system under study. The possibility to perform multi-speckle
experiments with a traditional DWS light scattering scheme thus
opens the pathway for a new type of fast and precise experiments.
If combined with the well established two-cell DWS technique
(TCDWS) \cite{scheffold01:dws} correlation times from 10 ns or
less up to duration of measurement can be accessed. Such improved
experimental performance is mandatory if progress shall be made in
the expanding field of slowly relaxing and arrested systems, such
as viscoelastic complex fluids, colloidal glasses and gels
\cite{cipelletti02:slow,pham02:multiple,eckert02:glass} .

%*********************************FIG1*********************************************

\begin{figure}
\includegraphics[width=0.6\linewidth]{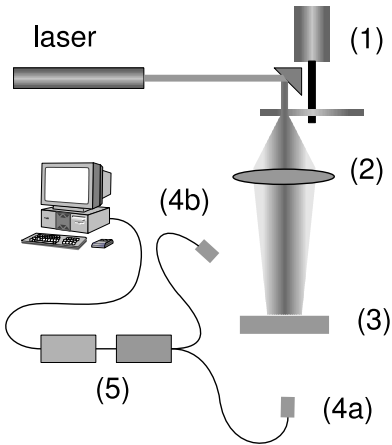}
\vspace{0.1cm} \caption{Experimental setup: Laser light is
scattered from a ground glass rotated by a fast stepper motor (1)
and the transmitted light is collected by a lens (2) to illuminate
the sample (3). Single mode fibers collect the scattered light
either in transmission (4a) or reflection (4b). The collected
light is subsequently analyzed by a single photon detector and
digital photon counter (5)} \label{fig:setup}
\end{figure}

%*********************************FIG1*********************************************

 Our experimental setup is shown in
Fig.~\ref{fig:setup}. A frequency-doubled Nd:YV$_{04}$ laser
("Verdi" from Coherent) operating at $\lambda_{0}=532$ nm is used
to illuminate a circular ground glass mounted on a 5-phases
stepper motor (RK-564 AC from VEXTA).  Through scattering and
dephasing the ground glass creates a speckle with a nearly
gaussian optical field \cite{basano:simple}. We collect the
transmitted light coming from the ground glass and focus it onto
the sample with a spot size diameter of roughly 5~mm. The
scattered light is then collected with a mono mode fiber and
analyzed by a photomultiplier and a digital photon counter
(correlator.com, New Jersey, USA). The photon counting device
records the time intervals between photons arrivals in a data file
with a resolution of $1/60$~MHz = 12.5 ns. The high temporal
resolution, comparable to the dead time of the detector, assures
that there is no more than one photon arriving at a time step for
a typical experimental count rate of 10-500 kHz. The stepper motor
is operated at frequencies up to 75 Hz. The measured intensity
correlation function (ICF) $[g^{(2)}_{M}(\tau)-1]$ contains
information from both the sample and the rotating diffuser. To
distinguish the dynamics due to the sample internal motion
$[g^{(2)}_{S}(\tau)-1]$ and due to the rotating diffuser
$[g^{(2)}_{S}(\tau)-1]$ in a quantitative way we take advantage of
previous studies of similar two-cell geometries. Scheffold et al.
have shown that the ICF $g^{(2)}_{M}(\tau)-1$ from a sandwich of
two optically independent cells can simply be expressed by a
product of the correlation functions of the two individual cells
\cite{scheffold01:dws}:
\begin{equation}\label{DoubleCell}
    g^{(2)}_{M}(\tau)-1=[g^{(2)}_{E}(\tau)-1]\cdot[g^{(2)}_{S}(\tau)-1],
\end{equation}

\begin{figure}
\centering
    \includegraphics[width=0.9\linewidth]{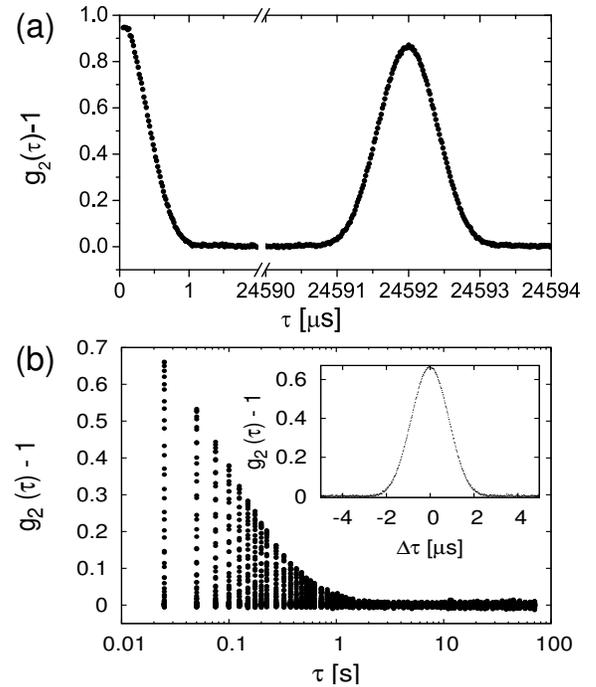}
\caption{ (a) 0 and 1st order correlation echo in
backscattering from a rigid teflon slab for a rotation frequency of
$f_r \approx 40$~Hz. The echo width at $1/e$ points is $2 \tau_r \approx 1.15$ \textmu s.
(b) DWS echoes in
backscattering for a sample of $TiO_2$ in glycerol. Inset shows the
echo shape for this sample.}
 \label{fig:echoes}
\end{figure}

 The two cell geometry can be also realized using a
very slowly rotating diffuser as suggested by Viasnoff et al.
\cite{viasnoff03:how}. In this version the two cells are separated
by a distance of several centimeters which ensures complete
decoupling of light propagation in both cells. This realization of
Two-Cell DWS is similar to our experimental setup. However,
in previous studies the diffuser was rotated slowly in order to average a
large number of arrested speckles over time whereas in our case
the diffuser motion is fast and periodical.
To illustrate the
different contributions we first consider the scattering signal from a rigid
sample without internal motion. Any fluctuation of the detected
laser light is produced in this case by the motion of the random diffuser.
Fig.~\ref{fig:echoes}(a) shows the typical intensity correlation function
of light scattered by a teflon slab of 2~mm thickness whith a motor rotation
frequency $f_{r}$ close to 40~Hz. At short times the rotation gives rise to a complete
decay of the ICF on a characteristic time $\tau_{r}$ set by $f_{r}$ and a corresponding echo width of $2\tau_{r}$. Each of the speckles reappear identical in the next
revolutions resulting in echoes in $[g^{(2)}_{E}(\tau)-1]$. At
$T_r=1/f_r$ and any multiple integer values $n=2,3,4,...$ a
correlation peak is observed. Echoes are found indistinguishable
for backscattering and transmission (data not shown). A detailed discussion of the
echo shape is beyond the scope of this letter but we expect many
similarities to the formalism developed for single scattering echoes
\cite{pham04:echo}. However, the number of speckles sweeping over the
detector can be readily estimated from our experiments to be $N
\approx T_{r} / 2\tau_{r}
> 2 \cdot 10^4$. This means already after $n + 1$
revolutions the correlation function of the $n$th echo is known
with an accuracy better than $1/\sqrt N \approx 1 \%$.

%*********************************FIG3*********************************************

%*********************************FIG3*********************************************
In the following sections we will discuss the application of the two cell
echo and compare the different data analysis schemes. We have
prepared a colloidal dispersion of titanium dioxide powder (Ref:
0041255 from Warner Jenkinson Europe Ltd.), particle diameter
roughly 200-300 nm, in glycerol at a volume fraction of $\approx
0.5 \%$. To further increase the viscosity the sample is kept at $5.7 \pm 0.5 ^\circ$ C . In this viscous opaque medium the correlation function
decays over a range of lag times accessible both to the echo
technique and traditional time averaging. Fig.~\ref{fig:echoes}
shows the result of the echo measurement in backscattering
geometry for a frequency of 40~Hz.

In the viscous glycerol solution the scatters undergo diffusive
motion expressed by the mean square particle displacement $\langle
r^{2}(\tau)\rangle =6D_{0}\tau$, where $D_{0}=k_{B}T/6\pi \eta a$
is the Einstein diffusion coefficient. For the backscattering
geometry the experimental ICF is described by the expression
\cite{maret1987}:

\begin{equation}
g_S^{(2)} (\tau ) - 1 = \beta  \cdot \exp \left[ { - 2\gamma \sqrt
{6\tau /\tau _0 } } \right]
    \label{eq:multi}
\end{equation}
with a relaxation time $\tau_0 = 1/D k{_0^2}$ and a factor $\beta$
that describes the intercept of the correlation function. $\gamma$
is a constant of order $\gamma \approx 2.1-2.3$ (VH
detection) \cite{rojasPRE2002:DWS}.

 It follows from Eq.~\eqref{eq:multi} that $g^{(2)}_{S}(\tau)-1$ can be extracted from the echo-peak
 value since at lag times $\tau=n/f_r$ one expects $g^{(2)}_{E}(n/f_r)-1 = 1$. However, the
echo peak hight might be affected by slight imperfections in the
rotation. Furthermore a detailed resolution of the peak maximum
can be costly in computation time in particular for higher
rotation frequencies. A more practical way of dealing with this
problem is to analyze the peak area rather than the peak height. Pham
et.al have shown for the case of single scattering echoes that the
peak area is directly proportional to ideal peak height, suffering
very little from slight imperfections in rotation
\cite{pham04:echo}. The peak aera can be obtained by numerical integration or simply by increasing the sampling
 time $\tau_s$. The latter approach is equivalent to triangular-weighted
integration \cite{schatzel:photon}  and moreover significantly reduces the
 computation time of $g_M^{(2)} (\tau ) - 1$.
\comment{
Choosing a lower resolution for the numerical
analysis of the ICF is furthermore equivalent to
triangular-weighted integration \cite{schatzel:photon} while at
the same time computation time is significantly reduced.
}

\begin{figure}
\centering
\includegraphics[width=\linewidth]{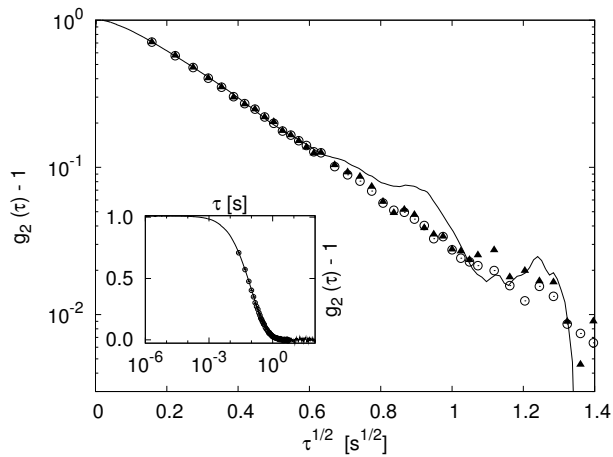}
\caption{Normalized ICF in backscattering (VH- geometry:
perpendicular polarization). Solid line - ICF from time averaging over 20 minutes, Symbols - echo analysis of a 12 second measurement
($\circ$) - data from time-of-arrival data processing,
($\blacktriangle$) - linear correlator with sampling time 12
\textmu s. Inset: Lin-log plot of the same data}
\label{fig:results}
\end{figure}

The relevant information is contained in the sample correlation
function $g_S^{(2)} (\tau ) - 1$. To study the influence of the
integration time window and the sampling time $\tau_s$ we have
varied both parameters over large range. Instead of numerical
integration we simply sum the points of the correlation function
$g^{(2)}_{M}(\tau)-1$. No significant dependence is found (within
reason) as long as the time window covers well the correlation
peak. As a matter of fact a single correlation channel of sampling
time $\tau_s=12$~\textmu s centered at the echo position provides
nearly the same level of accuracy as integration over 960 channel
at 12.5~ns resolution. In Fig.~\ref{fig:results} we compare the
echo-data collected during 12 seconds to a time averaged
measurement over 20 minutes. Dividing by the normalization factor
$\beta$ we obtain basically identical results from both
methods\footnote{With a glycerol ( 99\% ) viscosity of $\eta
\approx 5 Pa s$ at T=$5.7 ^\circ$C we expect a relaxation time
$\tau _0=1/D k_0^2 \approx 7.5-11.5$ seconds. From a fit of
equation (2) to the data we obtain $\tau _0$ = 8.1 seconds
($\gamma$=2.2)}. Despite a dramatically shorter measurement time,
the noise level is lower for the Echo-measurements.

In order to record these narrow correlation echoes special
emphasis has been given to the data analysis. In a traditional
linear correlator the ICF is obtained by averaging products of
photon counts in a certain sampling time interval $\tau_s$
separated by lag time $\tau$ (channel) \cite{berne76}. Over
several decades of lag times the linear channel layout however
becomes impractical and the computation time increases rapidly.
The usual solution to this problem in hardware correlators is to
use a multi-tau scheme as introduced by Sch\"atzel
\cite{schatzel:photon}: to decrease amount of data the sampling
time is doubled after a fixed number of steps. The correlation function is well resolved for the shortest lag times
while the resolution is decreased for longer lag times. This loss
in resolution is however not acceptable for our signal since all
echoes have the same width, even at large lag times. On the other
hand only a small number of correlation channels is needed to
resolve the ICF in the vicinity of the correlation echoes.
Furthermore not all echoes need to be resolved to cover a given
range of lag times. Thus, if the parameters are optimized, as shown
later in the text, simple multiplication can still be the most
efficient method.

There are alternative techniques that allow a fast computation of
the correlation function. One of the most efficient methods is
based on the fast Fourier transform (FFT). According to
the Wiener-Khinchin theorem the auto-correlation function of the
signal can be determined as an inverse Fourier transform of its
power spectrum  which is the square of
the Fourier amplitudes (for details see \cite{nr}). For a given sampling time $\tau_s$ FFT
provides the correlation function for all available channels  much
faster than any other technique. However, in order to to resolve the
echoes shape the sampling time has to be decreased
which requires large data arrays and long processing time.

Quite a different approach to calculate the ICF was put forward by
Chopra and Mandel \cite{chopra:correlator} in their photon
time-of-arrival correlator. It estimates the distribution of time
intervals between photons arrivals which is shown to be proportional
to the ICF. In contrast to the methods mentioned above computation can not be
accelerated by increasing the sampling time since photons arrivals have
to be recorded as separate events. This approach is much
faster only if high resolution data is required and if the data is
provided in a suitable format as it is the case for our
photon recorder. We have used this method to calculate the ICFs in
Fig.~\ref{fig:echoes} at the highest available resolution of
12.5~ns.

%*********************************FIG2*********************************************

%\begin{figure}[h]
%\centering
%\includegraphics[width=\linewidth]{TransBack_Static}
%\vspace{0.1cm} \caption{XXX order Correlation echoes in
%backscattering and transmission for a rotation frequency of XXX}
%\label{TransBack_Static}
%\end{figure}

%*********************************FIG2*********************************************

To optimize the data analysis we compare different methods: the
linear correlator, FFT and the time-of-arrival correlator. The
data of a 12 seconds measurement at an average count rate of 210 kHz
is analyzed under identical conditions. Correlation echoes are
calculated up to echo number 240, corresponding to $\tau$ = 6
seconds and the time delay between echoes is doubled after each
linear block of 16 and a total number of 60 echoes is
calculated. The time resolution of time-of-arrival correlator is
determined by the 12.5 ns hardware time step.
For the time of arrival-correlator the computation time is
ca. 14 seconds for an echo integration window of 6 \textmu based on a 1.7 GHz Intel Xeon processor system.
By calculating correlation coefficients only at the
echo peak position the processing time decreases to 10 seconds. The FFT
correlation function contains all available lag times and computation
time is inversely proportional to the sampling
time $\tau_s$. For the same data it takes 1.7 seconds with 12 $\mu$s
sampling time and less than one second for 24
$\mu$s. Processing the data with the linear correlator can be even faster. The whole
analysis takes approximately 1.5 seconds if a single channel for each
calculated echo with a sampling time of 12 \textmu s is choosen. Most of this time
is needed to convert the raw data to the format
\emph{photon-counts per sampling time}, which could be easily
done on-the-fly during data recording. In conclusion we find that the
time needed for an optimized data processing scheme is negligible compared to the total measurement
time.

%*********************************FIG4*********************************************

%*********************************FIG41*********************************************

Finally we would like to comment on the accessible time range at a
given rotation frequency.  The echo period can not be known with
absolute accuracy and at very high orders the sampling time
and echo period will not match any more and the signal is lost.
We commonly perform measurements up to echo number
1000 covering three orders of magnitude in lag-time. Since the echo width is only of the order of
1~\textmu s the period has to be known with nanosecond accuracy.
To overcome this difficulty for even higher order echos one might think of monitoring the echo period continuously
during processing. In this case, however, the echo shape has to be resolved in detail which goes at the expense of computation speed. A more practical way
to increase the time window is by performing several measurements at
different rotation speed. For such a scheme the integral time of
measurement will still be set approximately by the longest measurement.

In the summary we have shown that our Two-Cell Echo approach allows to
measure the ensemble averaged DWS correlation function nearly in
real time. Existing DWS experiments can be easily upgraded if the laser power is sufficient to drive the two-cell echo
experiment. Besides a simple device for precise mechanical
oscillation or rotation and a suitable photon counter or correlator no other
hardware is needed. Furthermore, combined with traditional
two-cell DWS at very low rotation speeds, the full intensity
autocorrelation function can be measured with a single
experimental setup covering more than 10 decades in correlation
time.

\acknowledgements Financial support by the Swiss National Science
Foundation, the TopNano21/KTI programme and the Marie Curie
network MRTN-CT2003-504712 is gratefully acknowledged. We thank
Suresh Bhat for help with the experiments and Peter Schurtenberger
for discussions.

%\include{echote}
%\bibliography{/usr/share/texmf/bibtex/bib/base/full}

\begin{thebibliography}{26}
\expandafter\ifx\csname natexlab\endcsname\relax\def\natexlab#1{#1}\fi
\expandafter\ifx\csname bibnamefont\endcsname\relax
  \def\bibnamefont#1{#1}\fi
\expandafter\ifx\csname bibfnamefont\endcsname\relax
  \def\bibfnamefont#1{#1}\fi
\expandafter\ifx\csname citenamefont\endcsname\relax
  \def\citenamefont#1{#1}\fi
\expandafter\ifx\csname url\endcsname\relax
  \def\url#1{\texttt{#1}}\fi
\expandafter\ifx\csname urlprefix\endcsname\relax\def\urlprefix{URL }\fi
\providecommand{\bibinfo}[2]{#2}
\providecommand{\eprint}[2][]{\url{#2}}

\bibitem[{\citenamefont{Cipelletti and Ramos}(2002)}]{cipelletti02:slow}
\bibinfo{author}{\bibfnamefont{K.A.}~\bibnamefont{Dawson}},  \bibinfo{journal}{Curr. Opin. Colloid Interface Sci.}
  \textbf{\bibinfo{volume}{7}}, \bibinfo{pages}{218}
  (\bibinfo{year}{2002}),
\bibinfo{author}{\bibfnamefont{L.}~\bibnamefont{Cipelletti}} \bibnamefont{and}
  \bibinfo{author}{\bibfnamefont{L.}~\bibnamefont{Ramos}},
  \bibinfo{journal}{ibid.}, \textbf{\bibinfo{volume}{7}}, \bibinfo{pages}{228} (\bibinfo{year}{2002}),\bibinfo{author}{\bibfnamefont{J.L.}~\bibnamefont{Harden}} \bibnamefont{and}
  \bibinfo{author}{\bibfnamefont{V.} \bibnamefont{Viasnoff}},
  \bibinfo{journal}{ibid.} \textbf{\bibinfo{volume}{6}},
  \bibinfo{pages}{438} (\bibinfo{year}{2001})

\bibitem[{\citenamefont{Pham et~al.}(2002)\citenamefont{Pham, Puertas,
  Bergenholtz, Egelhaaf, MoussaÎd, Pusey, Schofield, Cates, Fuchs, and
  Poon}}]{pham02:multiple}
\bibinfo{author}{\bibfnamefont{K.~N.} \bibnamefont{Pham}},
  \bibinfo{author}{\bibfnamefont{A.~M.} \bibnamefont{Puertas}},
  \bibinfo{author}{\bibfnamefont{J.}~\bibnamefont{Bergenholtz}},
  \bibinfo{author}{\bibfnamefont{S.~U.} \bibnamefont{Egelhaaf}},
  \bibinfo{author}{\bibfnamefont{A.}~\bibnamefont{MoussaÎd}},
  \bibinfo{author}{\bibfnamefont{P.~N.} \bibnamefont{Pusey}},
  \bibinfo{author}{\bibfnamefont{A.~B.} \bibnamefont{Schofield}},
  \bibinfo{author}{\bibfnamefont{M.~E.} \bibnamefont{Cates}},
  \bibinfo{author}{\bibfnamefont{M.}~\bibnamefont{Fuchs}}, \bibnamefont{and}
  \bibinfo{author}{\bibfnamefont{W.~C.~K.} \bibnamefont{Poon}},
  \bibinfo{journal}{Science} \textbf{\bibinfo{volume}{296}}
  (\bibinfo{year}{2002}).

\bibitem[{\citenamefont{Eckert and Bartsch}(2002)}]{eckert02:glass}
\bibinfo{author}{\bibfnamefont{T.}~\bibnamefont{Eckert}} \bibnamefont{and}
  \bibinfo{author}{\bibfnamefont{E.}~\bibnamefont{Bartsch}},
  \bibinfo{journal}{\pre} \textbf{\bibinfo{volume}{89}},
  \bibinfo{pages}{125701} (\bibinfo{year}{2002}).

\bibitem[{\citenamefont{Pusey and Megen}(1989)}]{pusey89:dls}
\bibinfo{author}{\bibfnamefont{P.}~\bibnamefont{Pusey}} \bibnamefont{and}
  \bibinfo{author}{\bibfnamefont{W.}~\bibnamefont{Megen}},
  \bibinfo{journal}{Physica A} \textbf{\bibinfo{volume}{57}},
  \bibinfo{pages}{705} (\bibinfo{year}{1989}).

\bibitem[{\citenamefont{Scheffold and
  Schurtenberger}(2003)}]{scheffold03:light}
\bibinfo{author}{\bibfnamefont{F.}~\bibnamefont{Scheffold}} \bibnamefont{and}
  \bibinfo{author}{\bibfnamefont{P.}~\bibnamefont{Schurtenberger}},
  \bibinfo{journal}{Soft Materials} \textbf{\bibinfo{volume}{1}},
  \bibinfo{pages}{139 } (\bibinfo{year}{2003}).

\bibitem[{\citenamefont{Xue et~al.}(1992)\citenamefont{Xue, Pine, Milner, Wu,
  and Chaikin}}]{xue92:nonergodicity}
\bibinfo{author}{\bibfnamefont{J.-Z.} \bibnamefont{Xue}},
  \bibinfo{author}{\bibfnamefont{D.}~\bibnamefont{Pine}},
  \bibinfo{author}{\bibfnamefont{S.}~\bibnamefont{Milner}},
  \bibinfo{author}{\bibfnamefont{X.}~\bibnamefont{Wu}}, \bibnamefont{and}
  \bibinfo{author}{\bibfnamefont{P.}~\bibnamefont{Chaikin}},
  \bibinfo{journal}{\pra} \textbf{\bibinfo{volume}{46}}, \bibinfo{pages}{6550}
  (\bibinfo{year}{1992}).

\bibitem[{\citenamefont{Sch\"atzel}(1993)}]{schatzel93:accuracy}
\bibinfo{author}{\bibfnamefont{K.}~\bibnamefont{Sch\"atzel}},
  \bibinfo{journal}{Applied Optics} \textbf{\bibinfo{volume}{32}},
  \bibinfo{pages}{3880} (\bibinfo{year}{1993}).

\bibitem[{\citenamefont{Beck et~al.}(1999)\citenamefont{Beck, Hartl, and
  Hempelmann}}]{beck99:glass}
\bibinfo{author}{\bibfnamefont{C.}~\bibnamefont{Beck}},
  \bibinfo{author}{\bibfnamefont{W.}~\bibnamefont{Hartl}}, \bibnamefont{and}
  \bibinfo{author}{\bibfnamefont{R.}~\bibnamefont{Hempelmann}},
  \bibinfo{journal}{The Journal of Chemical Physics}
  \textbf{\bibinfo{volume}{111}}, \bibinfo{pages}{8209} (\bibinfo{year}{1999}).

\bibitem[{\citenamefont{van Megen and Underwood}(1994)}]{megen94:glass}
\bibinfo{author}{\bibfnamefont{W.}~\bibnamefont{van Megen}} \bibnamefont{and}
  \bibinfo{author}{\bibfnamefont{S.~M.} \bibnamefont{Underwood}},
  \bibinfo{journal}{\pre} \textbf{\bibinfo{volume}{49}}, \bibinfo{pages}{4206 }
  (\bibinfo{year}{1994}).

\bibitem[{\citenamefont{Bartsch et~al.}(1997)\citenamefont{Bartsch, Frenz,
  Baschnagel, Sch\"artl, and Sillescu}}]{bartsch97:glass}
\bibinfo{author}{\bibfnamefont{E.}~\bibnamefont{Bartsch}},
  \bibinfo{author}{\bibfnamefont{V.}~\bibnamefont{Frenz}},
  \bibinfo{author}{\bibfnamefont{J.}~\bibnamefont{Baschnagel}},
  \bibinfo{author}{\bibfnamefont{W.}~\bibnamefont{Sch\"artl}},
  \bibnamefont{and} \bibinfo{author}{\bibfnamefont{H.}~\bibnamefont{Sillescu}},
  \bibinfo{journal}{\jcp} \textbf{\bibinfo{volume}{106}}, \bibinfo{pages}{3743}
  (\bibinfo{year}{1997}), \bibinfo{author}{\bibfnamefont{L.}~\bibnamefont{Cipelletti}} \bibnamefont{and}
  \bibinfo{author}{\bibfnamefont{D.}~\bibnamefont{Weitz}},
  \bibinfo{journal}{Rev. Sci. Instrum.} \textbf{\bibinfo{volume}{70}},
  \bibinfo{pages}{3214} (\bibinfo{year}{1999}), \bibinfo{author}{\bibfnamefont{A.}~\bibnamefont{Knaebel}},
  \bibinfo{author}{\bibfnamefont{M.}~\bibnamefont{Bellour}}, \bibinfo{author}{\bibfnamefont{J.-P.}~\bibnamefont{Munch}},
  \bibinfo{author}{\bibfnamefont{V.}~\bibnamefont{Viasnoff}},\bibinfo{author}{~\bibnamefont{F. Lequeux}} \bibnamefont{and}
  \bibinfo{author}{\bibfnamefont{J.L.}~\bibnamefont{ Harden}},
  \bibinfo{journal}{Europhys. Lett.} \textbf{\bibinfo{volume}{52}},
  \bibinfo{pages}{73} (\bibinfo{year}{2000}), \bibinfo{author}{\bibfnamefont{V.}~\bibnamefont{Viasnoff}},
  \bibinfo{author}{\bibfnamefont{F.}~\bibnamefont{Lequeux}} \bibnamefont{and}
  \bibinfo{author}{\bibfnamefont{D.}~\bibnamefont{Pine}},
  \bibinfo{journal}{Rev. Sci. Instrum.} \textbf{\bibinfo{volume}{73}},
  \bibinfo{pages}{2336} (\bibinfo{year}{2002}).



\bibitem[{\citenamefont{Berne and Pecora}(2000)}]{berne76}
\bibinfo{author}{\bibfnamefont{B.}~\bibnamefont{Berne}} \bibnamefont{and}
  \bibinfo{author}{\bibfnamefont{R.}~\bibnamefont{Pecora}},
  \emph{\bibinfo{title}{Dynamic Light Scattering. With Applications to
  Chemistry, Biology, and Physics}} (\bibinfo{publisher}{Dover Publications,
  Inc.}, \bibinfo{address}{New York}, \bibinfo{year}{2000}).

\bibitem[{\citenamefont{Maret and Wolf}(1987)}]{maret1987}
\bibinfo{author}{\bibfnamefont{G.}~\bibnamefont{Maret}} \bibnamefont{and}
  \bibinfo{author}{\bibfnamefont{P.-E.} \bibnamefont{Wolf}},
  \bibinfo{journal}{Z. Phys. B} \textbf{\bibinfo{volume}{65}},
  \bibinfo{pages}{409} (\bibinfo{year}{1987}).

\bibitem[{\citenamefont{Pine et~al.}(1988)\citenamefont{Pine, Weitz, Chaikin,
  and Herbolzheimer}}]{pine1988}
\bibinfo{author}{\bibfnamefont{D.}~\bibnamefont{Pine}},
  \bibinfo{author}{\bibfnamefont{D.}~\bibnamefont{Weitz}},
  \bibinfo{author}{\bibfnamefont{P.}~\bibnamefont{Chaikin}}, \bibnamefont{and}
  \bibinfo{author}{\bibfnamefont{E.}~\bibnamefont{Herbolzheimer}},
  \bibinfo{journal}{\prl} \textbf{\bibinfo{volume}{60}}, \bibinfo{pages}{1134}
  (\bibinfo{year}{1988}).


\bibitem[{\citenamefont{H\'ebraud et~al.}(1997)\citenamefont{H\'ebraud,
  Lequeux, and Munch}}]{hebraud97:yielding}
\bibinfo{author}{\bibfnamefont{P.}~\bibnamefont{H\'ebraud}},
  \bibinfo{author}{\bibfnamefont{F.}~\bibnamefont{Lequeux}}, \bibnamefont{and}
  \bibinfo{author}{\bibfnamefont{J.~P.} \bibnamefont{Munch}},
  \bibinfo{journal}{\prl} \textbf{\bibinfo{volume}{78}}, \bibinfo{pages}{4657 }
  (\bibinfo{year}{1997}).

\bibitem[{\citenamefont{H\"ohler et~al.}(1997)\citenamefont{H\"ohler,
  Cohen-Addad, and Hoballah}}]{hohler97:periodic}
\bibinfo{author}{\bibfnamefont{R.}~\bibnamefont{H\"ohler}},
  \bibinfo{author}{\bibfnamefont{S.}~\bibnamefont{Cohen-Addad}},
  \bibnamefont{and} \bibinfo{author}{\bibfnamefont{H.}~\bibnamefont{Hoballah}},
  \bibinfo{journal}{\prl} \textbf{\bibinfo{volume}{79}}, \bibinfo{pages}{1154 }
  (\bibinfo{year}{1997}).

\bibitem[{\citenamefont{van Megen et~al.}(1998)\citenamefont{van Megen,
  Mortensen, Williams, and M\"uller}}]{megen98:measurement}
\bibinfo{author}{\bibfnamefont{W.}~\bibnamefont{van Megen}},
  \bibinfo{author}{\bibfnamefont{T.~C.} \bibnamefont{Mortensen}},
  \bibinfo{author}{\bibfnamefont{S.~R.} \bibnamefont{Williams}},
  \bibnamefont{and} \bibinfo{author}{\bibfnamefont{J.}~\bibnamefont{M\"uller}},
  \bibinfo{journal}{\pre} \textbf{\bibinfo{volume}{58}}, \bibinfo{pages}{6073}
  (\bibinfo{year}{1998}).

\bibitem[{\citenamefont{Pham et~al.}(2004)\citenamefont{Pham, Egelhaaf,
  Moussa\"{i}, and Pusey}}]{pham04:echo}
\bibinfo{author}{\bibfnamefont{K.}~\bibnamefont{Pham}},
  \bibinfo{author}{\bibfnamefont{S.}~\bibnamefont{Egelhaaf}},
  \bibinfo{author}{\bibfnamefont{A.}~\bibnamefont{Moussa\"{i}}},
  \bibnamefont{and} \bibinfo{author}{\bibfnamefont{P.}~\bibnamefont{Pusey}},
  \bibinfo{journal}{Rev. Sci. Instrum.} \textbf{\bibinfo{volume}{75}},
  \bibinfo{pages}{2419 } (\bibinfo{year}{2004}).

\bibitem[{\citenamefont{Petekidis et~al.}(2002)\citenamefont{Petekidis,
  Moussaid, and Pusey}}]{petekidis02:rearrangements}
\bibinfo{author}{\bibfnamefont{G.}~\bibnamefont{Petekidis}},
  \bibinfo{author}{\bibfnamefont{A.}~\bibnamefont{Moussaid}}, \bibnamefont{and}
  \bibinfo{author}{\bibfnamefont{P.~N.} \bibnamefont{Pusey}},
  \bibinfo{journal}{\pre} \textbf{\bibinfo{volume}{66}}, \bibinfo{eid}{051402}
  (pages~\bibinfo{numpages}{13}) (\bibinfo{year}{2002}).

\bibitem[{\citenamefont{Basano and Ottonello}(1981)}]{basano:simple}
\bibinfo{author}{\bibfnamefont{L.}~\bibnamefont{Basano}} \bibnamefont{and}
  \bibinfo{author}{\bibfnamefont{P.}~\bibnamefont{Ottonello}},
  \bibinfo{journal}{J. Phys. E: Sci. Instrum.} \textbf{\bibinfo{volume}{14}},
  \bibinfo{pages}{1257 } (\bibinfo{year}{1981}).




\bibitem[{\citenamefont{Scheffold et~al.}(2001)\citenamefont{Scheffold,
  Skipetrov, Romer, and Schurtenberger}}]{scheffold01:dws}
\bibinfo{author}{\bibfnamefont{F.}~\bibnamefont{Scheffold}},
  \bibinfo{author}{\bibfnamefont{S.}~\bibnamefont{Skipetrov}},
  \bibinfo{author}{\bibfnamefont{S.}~\bibnamefont{Romer}}, \bibnamefont{and}
  \bibinfo{author}{\bibfnamefont{P.}~\bibnamefont{Schurtenberger}},
  \bibinfo{journal}{Phys. Rev. E} \textbf{\bibinfo{volume}{63}},
  \bibinfo{pages}{061404} (\bibinfo{year}{2001}).

\bibitem[{\citenamefont{Viasnoff et~al.}(2003)\citenamefont{Viasnoff, Jurine,
  and Lequeux}}]{viasnoff03:how}
\bibinfo{author}{\bibfnamefont{V.}~\bibnamefont{Viasnoff}},
  \bibinfo{author}{\bibfnamefont{S.}~\bibnamefont{Jurine}}, \bibnamefont{and}
  \bibinfo{author}{\bibfnamefont{F.}~\bibnamefont{Lequeux}},
  \bibinfo{journal}{Faraday Discussions} \textbf{\bibinfo{volume}{123}},
  \bibinfo{pages}{253 } (\bibinfo{year}{2003}).


\bibitem[{\citenamefont{Rojas et~al.}(2002)\citenamefont{Rojas-Ochoa,Romer,Scheffold,Skipetrov and Schurtenberger}}]{rojasPRE2002:DWS}
    \bibinfo{author}{\bibfnamefont{L.}~\bibnamefont{Rochas.Ochoa}},
  \bibinfo{author}{\bibfnamefont{S.}~\bibnamefont{Romer}},
\bibinfo{author}{\bibfnamefont{S.}~\bibnamefont{Skipetrov}},
\bibinfo{author}{\bibfnamefont{F.}~\bibnamefont{Scheffold}}
 \bibnamefont{and}
  \bibinfo{author}{\bibfnamefont{P.}~\bibnamefont{Schurtenberger}},
  \bibinfo{journal}{Phys. Rev. E} \textbf{\bibinfo{volume}{65}},
  \bibinfo{pages}{051403} (\bibinfo{year}{2002}).



\bibitem[{\citenamefont{Sch\"{a}tzel et~al.}(1988)\citenamefont{Sch\"{a}tzel,
  Drewel, and Stimac}}]{schatzel:photon}
\bibinfo{author}{\bibfnamefont{K.}~\bibnamefont{Sch\"{a}tzel}},
  \bibinfo{author}{\bibfnamefont{M.}~\bibnamefont{Drewel}}, \bibnamefont{and}
  \bibinfo{author}{\bibfnamefont{S.}~\bibnamefont{Stimac}},
  \bibinfo{journal}{J. Mod. Opt.} \textbf{\bibinfo{volume}{35}},
  \bibinfo{pages}{711} (\bibinfo{year}{1988}).

\bibitem[{\citenamefont{Press et~al.}(1992)\citenamefont{Press, Teukolsky,
  Vetterling, and Flannery}}]{nr}
\bibinfo{author}{\bibfnamefont{W.}~\bibnamefont{Press}},
  \bibinfo{author}{\bibfnamefont{S.}~\bibnamefont{Teukolsky}},
  \bibinfo{author}{\bibfnamefont{W.}~\bibnamefont{Vetterling}},
  \bibnamefont{and} \bibinfo{author}{\bibfnamefont{B.}~\bibnamefont{Flannery}},
  \emph{\bibinfo{title}{Numerical Recipes in C: The Art of Scientific
  Computing}} (\bibinfo{publisher}{Cambridge Univeristy Press},
  \bibinfo{address}{Cambridge}, \bibinfo{year}{1992}).

\bibitem[{\citenamefont{Chopra and Mandel}(1972)}]{chopra:correlator}
\bibinfo{author}{\bibfnamefont{S.}~\bibnamefont{Chopra}} \bibnamefont{and}
  \bibinfo{author}{\bibfnamefont{L.}~\bibnamefont{Mandel}},
  \bibinfo{journal}{Rev. Sci. Instrum.} \textbf{\bibinfo{volume}{43}},
  \bibinfo{pages}{1489 } (\bibinfo{year}{1972}).

%\bibitem{endnote27}{ With a glycerol ( 99\% ) viscosity of $\eta \approx 5 Pa s$ at T=$5.7 ^\circ$C we expect a relaxation time $\tau _0=1/D k_0^2 \approx 7.5-11.5$ seconds. From a fit of equation (2) to the data we obtain $\tau _0$ = 8.1 seconds ($\gamma$=2.2)}
\end{thebibliography}

\end{document}